\documentclass[letter]{aa}  

\usepackage{graphicx}
\usepackage{txfonts}
\usepackage{xcolor}
\begin{document}

\title{First high-resolution look at the quiet Sun with ALMA at 3 mm
}

\author{A. Nindos\inst{1} 
\and C.E. Alissandrakis\inst{1} 
\and T.S. Bastian\inst{2} 
\and S. Patsourakos\inst{1} 
\and B. De Pontieu\inst{3,4,5} 
\and H. Warren\inst{6} 
\and T. Ayres\inst{7} 
\and H.S. Hudson\inst{8,9}
\and T. Shimizu\inst{10}
\and J.-C. Vial\inst{11} 
\and S. Wedemeyer\inst{4}
\and V. Yurchyshyn\inst{12} }

\institute{Physics Department, University of Ioannina, Ioannina GR-45110, 
Greece\\
\email{anindos@uoi.gr}
\and
National Radio Astronomy Observatory, 520 Edgemont Road, Charlottesville VA 
22903, USA
\and
Lockheed Martin Solar and Astrophysics Laboratory, Palo Alto, CA 94304, USA
\and
Institute of Theoretical Astrophysics, University of Oslo, P.O. Box 1029 Blindern, N-0315 Oslo, Norway
\and
Rosseland Centre for Solar Physics, University of Oslo, P.O. Box 1029 Blindern, N-0315 Oslo, Norway
\and
Space Science Division, Naval Research Laboratory, Washington, DC 20375, USA 
\and
Center for Astrophysics and Space Astronomy, University of Colorado, Boulder
CO 80309, USA 
\and
School of Physics and Astronomy, University of Glasgow, G12 8QQ, Glasgow, UK
\and
Space Sciences Laboratory, University of California, 7 Gauss Way, Berkeley CA 
94720-7450,
USA
\and
Japan Aerospace Exploration 
Agency, 3-1-1 Yoshinodai, Chuo-ku, Sagamihara, Kanagawa 252-5210, Japan
\and
Institut d'Astrophysique Spatiale, UMR 8617, Universit\'{e} Paris 11, 91405, 
Orsay Cedex, France
\and
Big Bear Solar Observatory, New Jersey Institute of Technology, Big Bear City, 
CA 92314, USA\\
             }
\date{Received ...; accepted ...}

 
\abstract
{ 
We present an overview of high-resolution quiet Sun observations, from
disk center to the limb, obtained with the Atacama Large millimeter and 
sub-millimeter Array (ALMA) at 3 mm. Seven quiet-Sun regions were observed at a
resolution of up to 2.5\arcsec\ by 4.5\arcsec. We produced both
average and snapshot images by self-calibrating the ALMA visibilities
and combining the interferometric images with full-disk solar
images. The images show well the chromospheric network, which,
based on the unique segregation method we used, is brighter than
the average over the fields of view of the observed regions by $\sim
305$ K  while the intranetwork is less bright by $\sim 280$ K, with a
slight decrease of the network/intranetwork contrast toward the limb. 
At 3\,mm the network is very similar to the 1600\,\AA\ images, with somewhat 
larger size. We detect, for the first time, spicular structures, rising up to
15\arcsec\ above the limb with a width down to the image resolution
and brightness temperature of $\sim$ 1800 K above the local
background. No trace of spicules, either in emission or absorption,
is found on the  disk. Our results highlight the potential of ALMA for the
study of the quiet  chromosphere.
}

\keywords{Sun: radio radiation --
          Sun: chromosphere
               }

\titlerunning{Quiet-Sun observations with ALMA at 3 mm}
\authorrunning{Nindos et al.}

\maketitle
%


\section{Introduction}

The chromosphere, which is the layer from which most of the solar
radiation at millimeter (mm) wavelengths is emitted, is still not
fully understood.  Recent significant progress in modeling its
three-dimensional (3D) structure with sophisticated MHD models (see
Hansteen et  al. 2015 and references therein) has not succeeded in
capturing a  complete physical picture of the chromospheric fine
structure and its dynamics (e.g., see De Pontieu et  al. 2014, Tian et
al. 2014). The main obstacle is the complexity of practically all
available  diagnostics, apart from those in the radio spectrum.

Observations of the radio continuum at mm wavelengths provide a unique
chromospheric diagnostic. The quiet Sun mm-wavelength  emission
mechanism is free-free and electrons are almost always in local
thermodynamic equilibrium  (for details, see Shibasaki et al. 2011 and
Wedemeyer et al. 2016).

There have been numerous observations of the mm-$\lambda$ emission of the
Sun with single-dish telescopes; reviewed by Loukitcheva et
al. (2004). Most of them suffered from low spatial resolution and low
sensitivity. These observations have been used to constrain empirical
atmospheric models (see Shibasaki et al. 2011).

Interferometric observations of the Sun at mm wavelengths have been
few and most of them were performed with a small number of telescopes,
hence snapshot imaging was not possible (Labrum 1978; Belkora et
al. 1992; White and Kundu 1994). Probably the only exception in the
pre-Atacama Large millimeter and sub-millimeter Array (ALMA) era  is
the set of observations reported by White et al. (2006) with the
ten-element Berkeley-Illinois-Maryland Array (BIMA) at  85 GHz.

\begin{figure*}
\centering
\includegraphics[width=.86\textwidth]{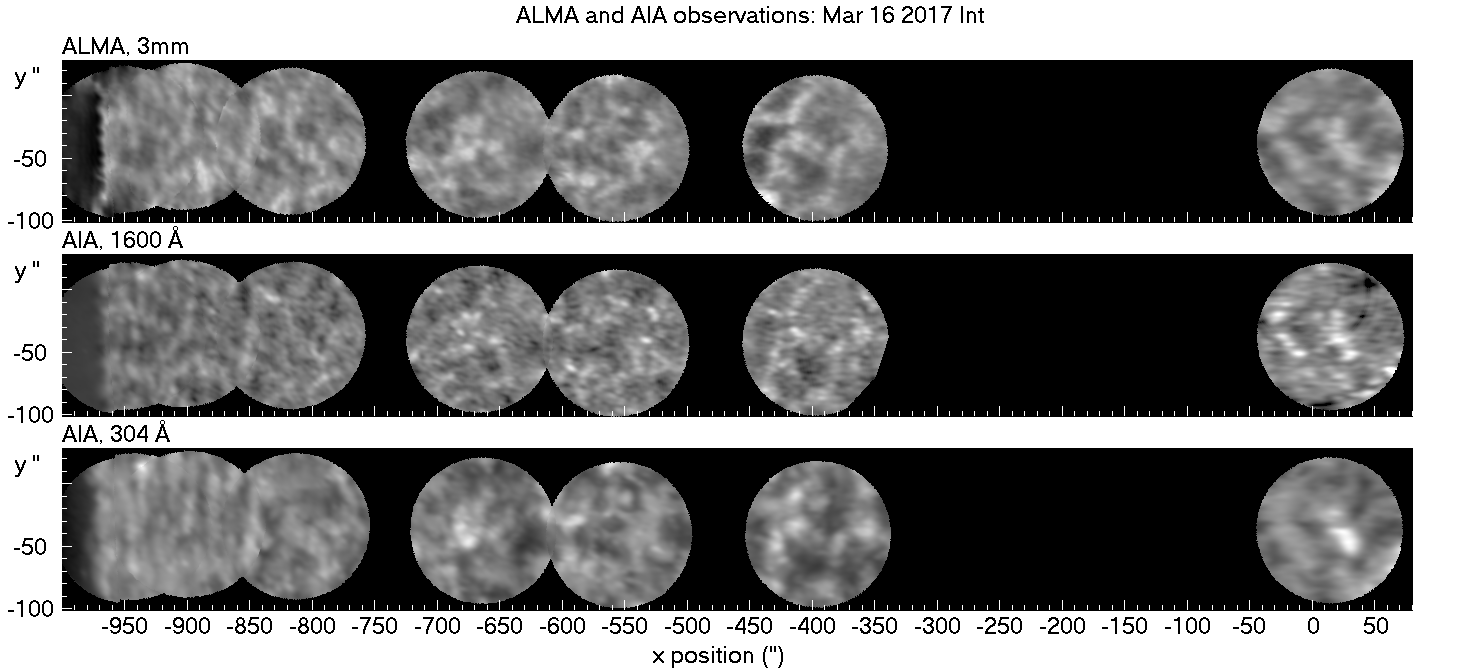}
\caption{Top row: Composite of the seven 3-mm average ALMA images, corrected for solar rotation and reprojected at the position of their FOVs at 17:55:48 UT. Second and third rows: Composites of AIA 1600 and  304\,\AA\ images for the same time intervals and FOVs as the ALMA images, convolved with the ALMA beam. ALMA and 1600\,\AA\ images are partly corrected for center-to-limb variation.  The composites are oriented in the SW direction with respect to solar north, $x$ and $y$ are with respect to the disk center. The photospheric radius is 964.8\arcsec. }
\label{TheWholeThing}
\end{figure*}

The availability of mm wavelength observations of the Sun with ALMA
provides the potential to advance our knowledge of the chromosphere
because of the instrument's unique capabilities in terms of spatial
resolution and sensitivity. Extensive commissioning activities  (see
Shimojo et al. 2017 and White et al. 2017) addressed the peculiarities
of solar observation with ALMA and culminated  in observations obtained
in  December 2015.  These data led to several publications,
reviewed by Loukitcheva (2018).  In those, there is only one
presentation of a quiet Sun image at 1\,mm  close to the limb (Shimojo
et al. 2017).

In a previous article (Alissandrakis et al. 2017; hereafter Paper I),
we used  ALMA full-disk (FD) commissioning data to study the quiet Sun
at low resolution; we computed the average center-to-limb variation
(CLV) of the brightness temperature, $T_b$, and we derived the
electron temperature, $T_e$, as a function of optical depth.  In this
article we present high-resolution quiet-Sun observations that we
obtained in March 2017; we present CLV measurements for the network
and the intranetwork and discuss the properties of spicules beyond the
limb.  


\section{Observations and data reduction}

Seven regions (targets) were observed at 100\,GHz on March 16, 2017,
from the limb to the center of the disk, along a line at a position
angle of 135\degr; each target was observed for 10\,min with a cadence
of 2\,s.  In order to have the best coverage of low spatial
frequencies, the  observations were performed in the most compact ALMA
configuration which included 40 twelve-meter antennas and 8 seven-meter antennas.
Twenty-two FD images were obtained  from two  12-m ``total power''
(TP) antennas (see White et al. 2017). The full width at half maximum (FWHM)
of the single-dish beam was about 60$''$ and each TP antenna  scanned
the full disk in 3 min. Further information about the
processing of the single-dish data is given in Appendix A.  The weather
conditions did not allow us to observe  at 239\,GHz, as we originally
intended.

Table \ref{ALMAobs} gives the middle of the observing  time
intervals, the pointings in solar coordinates (selected to have a good
sampling in $\mu$), the heliocentric distance, and $\mu$ for the seven
targets, as well as  the clean beam parameters (instrumental resolution). 
The resolution we achieved is close to the nominal resolution for the 
configuration we used which was $3.7\arcsec \times 2.5\arcsec$.
The positions of the targets on a FD image are shown in Fig. \ref{FD}.
 
\begin{table}[h]
\begin{center}
\caption{Pointing and clean beam size of the ALMA 3\,mm observations.}
\label{ALMAobs}
\begin{tabular}{lccccccccc}
\hline 
Tgt& Aver UT&     x     &     y    &R/R$_\odot$ &$\mu$ & $B_{min}$ &$B_{maj}$ & P.A.   \\
   &        & (\arcsec) & (\arcsec)&            &      & (\arcsec) &(\arcsec) & (\degr)\\
\hline
 1  &15:27:35& $-$680 & $-$680 & 0.97 & 0.16 & 2.4 & 4.5 & $-$84 \\
 2  &16:19:51& $-$650 & $-$654 & 0.94 & 0.34 & 2.4 & 4.5 & $-$82 \\
 3  &17:01:38& $-$585 & $-$585 & 0.85 & 0.52 & 2.6 & 4.6 & $-$79 \\
 4  &18:04:06& $-$488 & $-$464 & 0.69 & 0.72 & 2.5 & 4.8 & $-$64 \\
 5  &18:45:54& $-$377 & $-$404 & 0.56 & 0.82 & 2.5 & 5.0 & $-$59 \\
 6  &19:28:03& $-$265 & $-$289 & 0.40 & 0.92 & 2.5 & 5.5 & $-$52 \\
 7  &20:23:29&   ~~50 &   ~~~0 & 0.05 & 1.00 & 2.3 & 8.1 & $-$48 \\
\hline 
\end{tabular}
\end{center}
\end{table}

The visibilities were calibrated following the method described by
Shimojo et al. (2017) and were provided to us by the Joint ALMA
Observatory together  with average CLEAN images. These images were not
corrected for the primary beam response and no FD component had
been added.  Comparison of these images with simultaneous 1600
\AA\ images obtained by the Atmospheric Imaging Assembly (AIA) aboard
Solar Dynamics Observatory (SDO) revealed that the ALMA pointing
information provided in the visibilities  was not accurate. Therefore
the actual pointing was computed by cross-correlation of the ALMA
with the AIA 1600 \AA\ images.
 
We imported the visibilities into the Astronomical Image Processing
System (AIPS) and for each target we re-calculated CLEAN images
synthesized from the entire observing period. We also calculated 2\,s
snapshot images. Both our average and snapshot CLEAN images were
produced after  four rounds of phase self-calibration, with solution
intervals of 20 min, 1 min, 2 s, and 2 s, and one round of amplitude
self-calibration with a solution interval of 2 s. The self-calibration,
which is not included in the standard data pipelines, improved the
image quality and removed the jitter that was apparent in the movies
of snapshot images computed from the visibilities after the standard
calibration procedure. 

Subsequently, all images were imported into the Common Astronomy
Software Applications (CASA) software, where they were combined with
the processed FD image that is present in Fig. A.1.  This procedure is
necessary to (1) recover low spatial frequencies not accessible to the
interferometric observations, and (2) obtain absolute brightness
temperatures from the ALMA data.  The combination was done using
CASA's task \texttt{feather} and correcting for the fact that the
flux of the derived images was higher than that of the FD image by
14\%,  as noted in Paper I.  Beyond the limb, we noticed in Target 1
an artificial $T_b$ enhancement as high as 1000 K (see Fig. 2).  This
artifact may result from the incomplete sampling of the u-v plane
and/or  non-optimal weighting of the visibilities (see Shimojo et
al. 2017 for details). The last step in the data reduction was the
correction of all images for the primary beam response.

\section{Results}

A composite of all average ALMA images, each with a field of view (FOV) of 120\arcsec\ is presented in the top row of Fig. \ref{TheWholeThing}; the other rows show similar composites of the corresponding AIA data in the 1600 and 304\,\AA\ bands, convolved with the ALMA beam. In the course of this comparison we found a displacement towards the limb of the 304 images with respect to those at 1600, indicating that the 304\,\AA\ emission forms 4.8\arcsec\  (3\,Mm) higher  and this effect has been taken into account in Fig. \ref{TheWholeThing}.

A first remark is that the quality of the ALMA images is very uniform
over the FOV, which is twice the nominal FOV of 60\arcsec, (defined as
the FWHM of a single antenna), with some exceptions near the edge.
The chromospheric network is clearly visible at 3 mm, its morphology
being very similar to that in the two AIA bands.  There is also a very
strong similarity between the ALMA images and the Interface Region Imaging
Spectrograph (IRIS) slit-jaw images in three bands: 2796 (Mg\,{\sc ii}), 1440 
(Si\,{\sc iv}) and 1330\,\AA\ (C\, {\sc ii}).

Using full-disk ALMA images, we noted in Paper I that the 239\,GHz network correlates slightly better with 1600\,\AA. Here with high resolution we find that
the 100\,GHz images correlate equally well with the 1600\,\AA\ and the 304\,\AA\ images (cross-correlation of $\sim0.7$). Moreover, at 100\,GHz the structures are larger than at 1600\,\AA, but not as large as at 304\,\AA\
(for targets 3-7 the width of the autocorrelation function is $\sim$25\%
narrower at 1600 and $\sim$30\% wider at 304 \AA)
suggesting that the radiation is formed in between. We note that at cm-$\lambda$ the network correlates best with the 304\,\AA\ band \citep{2015SoPh..290....7B}.

\begin{figure}[h]
\centering
\includegraphics[width=9cm]{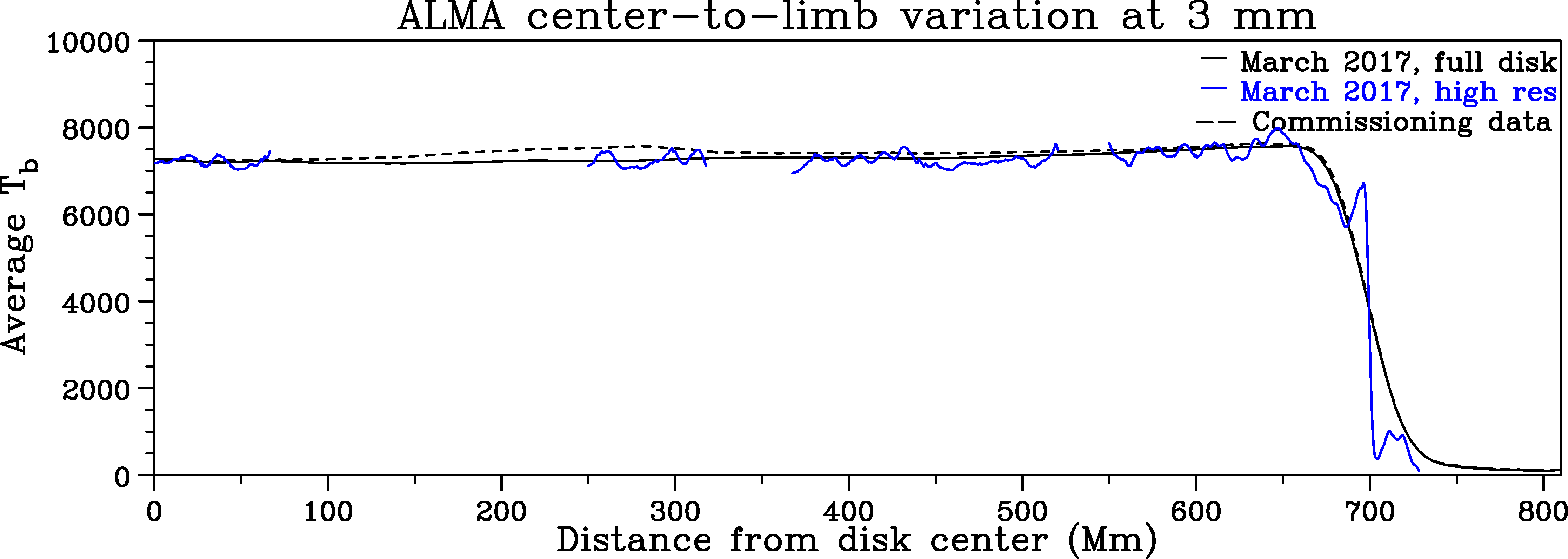}
\caption{Center-to-limb variation from the present and the commissioning full-disk data (full and dashed black curves) and from the interferometric average images with a FOV of 110\arcsec\ (blue curve).}
\label{clv}
\end{figure}

Figure \ref{clv} shows the average brightness temperature, $T_b$, derived from the ALMA interferometric images as a function of distance from disk center
by performing azimuthal averaging. In the same figure we give the corresponding curves from the FD images of the present and the commissioning data (Fig. 4 of Paper I). We first note that the FD curves of this work and Paper I are remarkably similar, apart from some disk structures in the latter; indeed, during the present observations the Sun was much more quiet than in December 2015 (cf. Fig. 1 of Paper I with Fig. \ref{FD}). Apart from that, the high-resolution data of Fig. \ref{clv} show a moderate limb brightening, $\sim5$ \arcsec\ wide, just beyond the photospheric limb. 

\begin{table}[h]
\begin{center}
\caption{Network, cell interior average and RMS $T_b$  in K.}
\label{stats}
\begin{tabular}{cccccccccc}
\hline 
$\mu$ & Net & Cell & Aver & RMS &Net/Cell & Net\,$-$ & Net\,$-$ & Cell\,$-$ \\
      &     &      &      &     &Ratio    &   Cell   &   Aver   &    Aver   \\
\hline                                                           
0.34 & 7670 & 7120 & 7370 & 570 & 1.077 & 550 & 300 & $-$250 \\
0.52 & 7760 & 7150 & 7460 & 440 & 1.084 & 600 & 300 & $-$310 \\
0.72 & 7500 & 7030 & 7270 & 400 & 1.066 & 470 & 240 & $-$230 \\
0.82 & 7670 & 7030 & 7330 & 430 & 1.091 & 640 & 340 & $-$300 \\
0.92 & 7620 & 6950 & 7270 & 460 & 1.097 & 670 & 350 & $-$330 \\
1.00 & 7530 & 6940 & 7220 & 390 & 1.085 & 590 & 300 & $-$290 \\
\hline 
\end{tabular}
\end{center}
\end{table}

\begin{figure*}
\centering
\includegraphics[width=.76\textwidth]{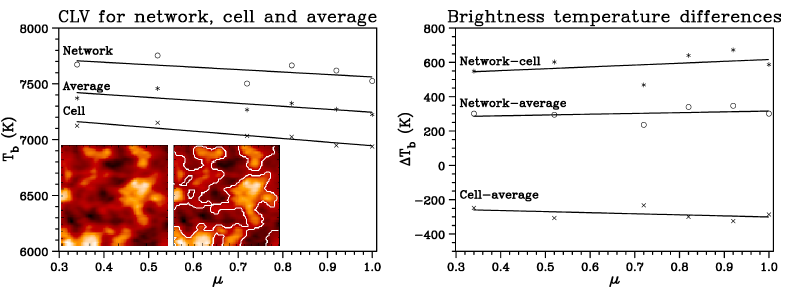}
\caption{$T_b$ of the network, cell interior, and average (left); and their differences (right) as a functon of $\mu$. The insert shows the inner $80\arcsec \times80$\arcsec\ of Target 5; contours show the network-cell boundaries.}
\label{statsfig}
\end{figure*}

In order to quantify the center-to-limb variation (CLV) of the network and cell interior we computed the average $T_b$ of both for Targets 2-7, as well as the average $T_b$ and its root mean square (RMS) value. We limited the FOV to 80\arcsec\ to minimize edge effects. We next fitted $T_b(x,y)$ with a second-degree polynomial to avoid large-scale structure and we considered the average of the values above the fit as representative of the network and the others as representative of the intranetwork. 
This method of separating network/internetwork is by no means unique, but leads to a reasonably good network/internetwork segregation (see insert in Fig. \ref{statsfig}).
The results are given in Table \ref{stats} and plotted in Fig. \ref{statsfig}. 

Both the network and the cell interior show limb brightening. Their difference, as well as the difference between the network and the average $T_b$, decrease marginally towards the limb, while the difference between the cell interior and average difference shows a marginal increase. The network/cell ratio has an average value of 1.08 and decreases slightly towards the limb. On average, the network is $305 \pm 35$\,K above the average $T_b$ and the intranetwork $285 \pm 30$\,K below. Our results are only slightly affected by the ALMA resolution; 
a check using AIA 1600\,\AA\ data for Target 5 gave a 3.3\% increase of the cell intensity and a 3.7\% drop of the network intensity after convolution with the ALMA beam.

\begin{figure*}
\centering
\includegraphics[width=.76\textwidth]{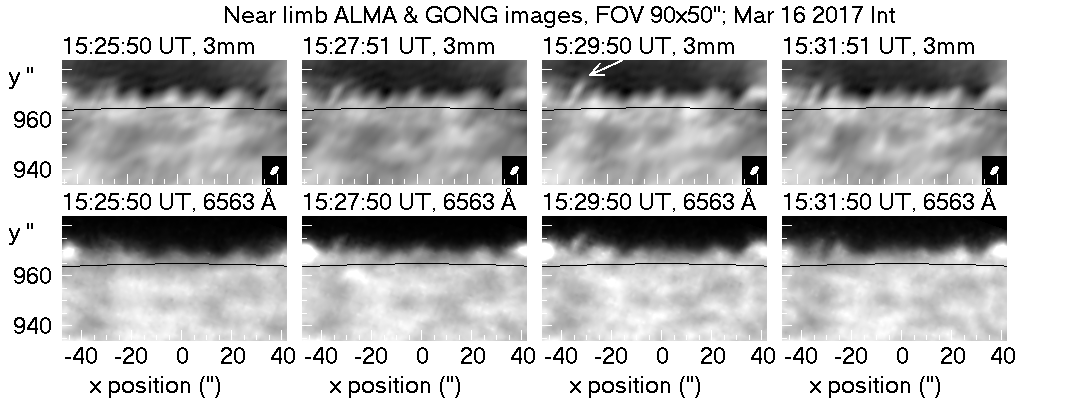}
\caption{Spicules seen in near-limb images at 3 mm (top) and H$\alpha$ (bottom). All images have been rotated to make the limb parallel to the horizontal direction and corrected to reduce the sharp drop of intensity near the limb
by subtracting 90\% of the azimuthally averaged intensity.
The black curve marks the photospheric limb and the inserts show the ALMA beam.}
\label{spicules}
\end{figure*}

Structures beyond the limb are easily visible in Fig. \ref{TheWholeThing}. A closer look shows that they are spicules, or rather groups of spicules (Fig. \ref{spicules}), as evidenced from the comparison of ALMA with H$\alpha$ images from the GONG network. The most prominent structure, on the left and marked by an arrow 
in Fig. 4, is almost parallel to the ALMA beam, thus facilitating its detection. It is visible from $\sim2$ to $\sim15$\arcsec\ above the limb and is most prominent at 8.5\arcsec, with $T_b=2560$\,K, 1800\,K above the local background and a width of 4\arcsec. Lower down, $T_b$ goes up to 6500\,K and the width drops to 2.5\arcsec, very close to the instrumental resolution of 2.4\arcsec. Their lifetime is comparable to the 10\,min duration of the Target 1 observations. Although the roots of some spicular structures in Fig. \ref{spicules} are inside the limb, we found no evidence of spicules on the disk, either in emission or in absorption.

\section{Summary and discussion}
Although the weather conditions during our ALMA observations were far from ideal, we obtained good average and snapshot images through the tedious process of self-calibration, with a resolution very close to the theoretical ($2.4\arcsec \times 4.5$\arcsec). 
The most serious problem with our final images is that we got emission beyond the limb of up to 1000\,K (14\% of the disk center value), the origin of which we could not identify. This excess emission occurs well beyond the region where the spicules are located and therefore its presence does not affect the brightness temperature measurements of the spicules. 
Apart from that, our images appear good over a circular FOV with a diameter of about 120\arcsec, which is twice the nominal ALMA FOV at 100\,GHz. The good quality of the images was judged by the fact that no significant artifacts appeared close to the edge of the FOV.

Our images show clearly the chromospheric network; morphologically
they are close to AIA 1600\,\AA\ and 304\,\AA.  Individual ALMA
features are broader than at 1600\,\AA, indicating that the 3\,mm
radiation on the disk is formed between the 1600 and the 304 levels;
we found that the latter is formed  3\,Mm above the 1600 level. 

The CLV curve from the present FD data is remarkably similar to the
one we derived from the commissioning data in Paper I, while the
high-resolution data show a moderate and narrow limb brightening just
above the photospheric limb. The network is brighter than the average
by $\sim305$\,K and the intranetwork is less bright by $\sim280$\,K,
with a slight decrease of the network/intranetwork contrast towards the 
limb; this is crucial information for atmospheric modeling. 

We report for the first time the detection of spicular structure
beyond the limb, with a width down to our theoretical resolution,
detectable up to 15\arcsec\ above the limb and with $T_b\sim1800$\,K
above the local background.  Interestingly, no trace of spicules
appears in the 1-mm image in Shimojo et al. (2017). We see no trace of
spicules on the disk, either in emission or in absorption, probably
due to insufficient resolution or small optical thickness; further
work is required to deduce whether their effect on the CLV of the
brightness is small or, as is usually assumed (e.g., Lantos and Kundu
1972; Selhorst et al. 2005), substantial.

In this article we present a broad overview of the first high-resolution ALMA observations of the network at 3 mm. In subsequent work we intend to exploit the CLV data for two-component atmospheric modeling, elaborate further on the larger size of the structures at 3\,mm than at lower heights, compare with magnetograms, and study oscillations and transients
and their effect on the corona. From the observational point of view, similar data at 239\,GHz and higher frequencies are highly desirable, as well as measurements of circular polarization and observations with more extended configurations and thus higher resolution.  

\begin{acknowledgements}
This paper makes use of the following ALMA data:
ADS/JAO.ALMA\#2016.1.00572.S. ALMA is a partnership of ESO
(representing its member states), NSF (USA) and NINS (Japan), together
with NRC (Canada) and NSC and ASIAA (Taiwan), and KASI (Republic of
Korea), in cooperation with the Republic of Chile. The Joint ALMA
Observatory is operated by ESO, AUI/NRAO and NAOJ.
IRIS  is  a  NASA small explorer mission developed and operated by LMSAL with mission operations executed at NASA Ames Research center and major contributions to downlink communications funded by ESA and the Norwegian Space Centre.
\end{acknowledgements}

%
%

\begin{appendix}

\section{Full-disk images}

\begin{figure}[h]
\centering
\includegraphics[width=\hsize]{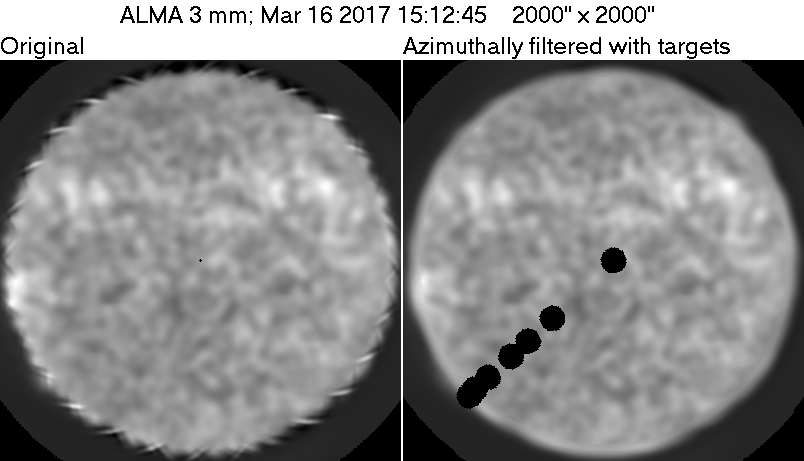}
\caption{Full-disk ALMA images: Left: Original. Right: After azimuthal filtering, correction of the limb ellipticity and with the seven targets added as black disks of 120\arcsec\ diameter. In both images 90\% of the azimuthally averaged intensity has been subtracted to enhance the visibility of disk features.}
\label{FD}
\end{figure}

During our ALMA observation, 22 FD images were obtained
with antennas 0 and 1. Figure \ref{FD} (left) shows one of them, partly
corrected for center-to-limb variation. As in our previous work
(cf. Fig. 1 of Paper I), the FD image shows spoke-wheel-like
irregularities near the limb, which could influence the ``feathering''
procedure for near-limb Target 1. We reduced these features by
low pass azimuthal filtering of the region near the limb with a
120\arcsec\ wide Gaussian. We also corrected the FD image for the
slight ellipticity of the solar disk, also reported in Paper I. The corrected image is shown at the right of
Fig. \ref{FD}; the dark disks mark the position of the seven
targets and their diameter is equal to a FOV of 120\arcsec.

In order to facilitate comparison of future modeling studies of our
data with those presented in Paper I, we scaled the FD image to match
the quiet Sun brightness temperature at disk center reported in Paper
I which was 7250 K. This value is only 50 K lower (about 0.7\%) than the value
recommended by White et al. (2017), that is, 2 and 3.4 times smaller than
the relevant uncertainties reported in  White et al. (2017) and Paper
I, respectively.

We found that the addition of the large-scale spatial components did not change the image structure appreciably, except for the near-limb Target 1. For this target we took particular care to match the FD and interferometric limbs, by matching the inflection points of their CLV curves. 
 
\end{appendix}

\end{document}